\documentclass[a4paper,nobibnotes,nofootinbib]{revtex4}

\usepackage{amsmath,amssymb}
\usepackage{epsfig}

\newcommand{\ccc}{\chi^{\prime\prime}(\gamma_c)}

\begin{document}
\title{Traveling wave fronts and the transition to saturation}

\author{S. Munier}\email{Stephane.Munier@cpht.polytechnique.fr} 
\affiliation{Centre de physique th{\'e}orique,
{\'E}cole polytechnique, 91128 Palaiseau cedex, France%
\footnote{UMR 7644, unit{\'e} mixte de recherche du CNRS.}}
\author{R. Peschanski}
\email{pesch@spht.saclay.cea.fr}
\affiliation{Service de physique th{\'e}orique, CEA/Saclay,
  91191 Gif-sur-Yvette cedex, France\footnote{%
URA 2306, unit{\'e} de recherche associ{\'e}e au CNRS.}}

\begin{abstract}
We propose a general method to study the solutions to nonlinear QCD
evolution equations, based on a deep analogy with the physics of
traveling waves. In particular,
we show that the transition to the saturation regime 
of high energy QCD
is identical to the formation of the front of a traveling wave. 
Within this physical picture,
we provide the expressions for the saturation scale and the gluon
density profile as a function of
the total rapidity and the transverse momentum. 
The application to the Balitsky-Kovchegov equation for both fixed and
running coupling constants confirms the effectiveness of this method.
\end{abstract}

\maketitle

\section{Introduction}

Phenomenology of high energy deep-inelastic scattering in the 
framework of QCD leads to the study of
evolution equations
for the gluon distribution function ${\cal N}(k,Y)$
at high density
\cite{GLR,qiu,venugopalan,Balitsky:1995ub,levin,Mueller:2002zm}.
$Y$ is the rapidity of the evolved gluons and plays the r{\^o}le of 
the evolution variable, and
$k$ is
their transverse momentum.
The main feature of these evolution equations is that they contain
nonlinear damping terms which make them difficult to solve by general methods.
These damping terms reflect the effect of saturation, which is due to the
recombination of partons densely packed together.
The evolution at low density is well-understood and is described by a
linear equation. The deep saturation regime can also be evaluated.
However, the transition
between these two regimes
is still a challenge. Some interesting approaches for  solving  non-linear 
evolution equations have been proposed \cite{GLR,levin}. Recently, the use of 
solutions of  the linear evolution supplemented by ``absorbing'' boundary 
conditions has been proposed to determine the energy dependence of the 
saturation momentum \cite{Mueller:2002zm}.

The purpose of the present paper is to propose a new method
for handling this problem. It is based on a
deep connection that we have found
between the formation of traveling
wave fronts in nonlinear physics and the transition to saturation.

We expect our method to be very general and applicable for various
types of equations.
In the present paper, we are going to apply it
in detail to the 
Balitsky-Kovchegov
(BK) equation \cite{Balitsky:1995ub}, which reads
\begin{equation}
{\partial_Y}{\cal N}=\bar\alpha
\chi\left(-\partial_L\right){\cal N}
-\bar\alpha\, {\cal N}^2\ ,
\label{bk}
\end{equation}
where $L=\log (k^2/\Lambda_{\mbox{\footnotesize QCD}} ^2)$.
The QCD coupling $\bar\alpha$ is either fixed or runs with $k$, in which
case it is given by
\begin{equation}
\bar\alpha(L)=\frac{\alpha_s(L)N_c}{\pi}=\frac{1}{bL}\ ,\ \ \
b=\frac{11N_c-2N_f}{12N_c}\ .
\label{baralpharun}
\end{equation}
The function $\chi$ is the well-known 
Balitsky-Fadin-Kuraev-Lipatov
(BFKL) kernel \cite{BFKL}
\begin{equation}
\chi(\gamma)=2\psi(1)-\psi(\gamma)-\psi(1-\gamma)
\label{chi}
\end{equation}
and $\chi\left(-\partial_L\right)$ is an integro-differential
operator which may be defined 
with the help of the formal series expansion
\begin{equation}
\chi\left(-\partial_L\right)=
\chi(\gamma_0){\bf 1}+\chi ^\prime(\gamma_0)(-\partial_L-\gamma_0{\bf 1})
+{\scriptstyle\frac12}\chi ^{\prime\prime}(\gamma_0)
(-\partial_L-\gamma_0{\bf 1})^2
+{\scriptstyle\frac16}\chi ^{(3)}(\gamma_0)
(-\partial_L-\gamma_0{\bf 1})^3+\cdots
\label{chiexpansion}
\end{equation}
for some given $\gamma_0$ between 0 and 1, {\it i.e.} for the principal branch
of the function $\chi$.
The BK equation is 
expected to capture the main features of saturation effects
in high energy scattering processes like virtual
photon-nucleus (or proton under certain conditions) total cross
section. 

In a previous paper and in the case of fixed QCD coupling, 
we have already shown \cite{Munier:2003vc}
the intimate connection between the 
limiting speed of a wave front and geometric scaling, namely the dependence
of ${\cal N}(k,Y)$ on a unique scaling variable $k/Q_s(Y)$, where
$Q_s$ is the so-called {\it saturation scale}.
Interestingly enough, geometric scaling has been observed in 
the HERA data on inclusive $\gamma ^*\!-\!p$ scattering
\cite{Stasto:2000er}.
One aspect of the present letter is to go beyond these asymptotic
results and to show how the approach to
geometric scaling is related to
the mechanism of wave front formation.
This requires a method to get information on the
form of the wave front when rapidity increases.
Another goal is to extend our analysis to both fixed and running
QCD coupling.

In Sec.\ref{2}, we present the general method that we propose.
In Sec.\ref{3}, we treat the fixed coupling problem and
in Sec.\ref{4} the running coupling 
one. For both cases
we derive the expressions for 
the saturation scale $Q_s(Y)$ 
and for the gluon density
${\cal N}(k/Q_s(Y),Y)$
in the vicinity of 
the geometric scaling
region $L\sim\log(Q_s^2/\Lambda^2_{\text{\footnotesize QCD}})$.
The final Sec.\ref{5} is devoted to a discussion and summary.


\section{Phase velocity, group velocity and wave front formation}
\label{2}

Let us
outline the method how to investigate 
traveling wave solutions to nonlinear partial
derivative equations.

Our analysis comes from the physics of 
linearly unstable front equations of diffusion type. They appear {\it
  e.g.} for Rayleigh instabilities in hydrodynamical flows, directed
polymers in random media, chemical and bacterial growth models
etc...\cite{vansaarloos}.
The evolution equations have the property to
admit traveling wave solutions. 
Their existence is due to nonlinear damping while
some of their large time characteristic features
(front velocity, front shape) are determined by the {\it linearized}
equation about the unstable state.
The front is said to be ``pulled'' by this unstable state which appears
as an unstable fixed point of the equation.
The main r{\^o}le of the nonlinearities 
is to cause saturation.

Let us illustrate these features on the generic equation
\begin{equation}
\partial_t u(x,t)={\mathbb D}\cdot u(x,t)+f(u)
\label{general}
\end{equation}
where ${\mathbb D}$ is a linear operator acting on functions of $x$, 
$f(u)$ contains the nonlinearities, with the requirement that
${\mathbb D}\cdot u(x\!=\!\mbox{const.},t)=0$, and $f(0)=0$, 
so that $u=0$ is a fixed
point. This fixed point is furthermore required to be unstable
with respect to perturbations, which has to be checked in each
particular case.

The simplest case is the Fisher and 
Kolmogorov-Petrovsky-Piscounov (KPP)
equation \cite{Fisher:37}
corresponding to ${\mathbb D}=\partial^2_x$ and
$f(u)=u-u ^2$, for which rigorous mathematical results are available
\cite{Bramson}.
Using a change of
variables, we have shown \cite{Munier:2003vc} that 
the BK equation reduces to the KPP equation when its
kernel~(\ref{chi}) 
is approximated by the first three terms of the 
expansion~(\ref{chiexpansion}).

More generally, on physical grounds\footnote{%
At variance with the case of the KPP equation, mathematical
proofs are lacking for properties of general
nonlinear equations. However,
a large number of results \cite{vansaarloos} 
have been found
which
{\it e.g.} numerically confirm the existence and properties of
traveling wave solutions to equations of the 
type~(\ref{general}).}, we
expect the methods developed to treat the KPP equation to be valid
for any saturation equation in QCD.
Indeed, the common feature of these equations
is that their nonlinearities
cause saturation of the parton densities, 
and the fixed point of low gluon
density is clearly linearly unstable under rapidity evolution.
This instability is physically related
to the growth of parton densities with rapidity given by the BFKL
equation \cite{BFKL}.

We shall study the form of the wave front 
at large times
by working in its own reference
frame which is defined by 
the change of variable
$x=x_{\text{\tiny WF}}+vt$, and in the neighbourhood of the front, namely
$x_{\text{\tiny WF}}\ll vt$. The velocity $v$ of the wave front will be
determined by consistency as the velocity of
the slowest wave present in the initial wave packet.

The solution to the linear part of equation~(\ref{general}) 
corresponds to a linear superposition of waves:
\begin{equation}
u(x,t)=\int_{\cal C}
\frac{d\gamma}{2i\pi}\,u_0(\gamma)\,
\exp \left\{{-\gamma (x_{\text{\tiny WF}}+vt)
+\omega(\gamma)t}\right\}\ ,
\label{inicond}
\end{equation}
where $\omega(\gamma)$ is the Mellin transform of the linear kernel 
($\partial_x^2+{\bf 1}$ for the KPP equation) and 
defines the dispersion relation of the linearized equation.
In particular, each partial wave of wave number
$\gamma$ has a {\it phase velocity} 
\begin{equation}
v_{\varphi}(\gamma)=\frac{\omega(\gamma)}{\gamma}
\end{equation} 
whose expression is found by imposing
that the exponential factor in 
Eq.(\ref{inicond}) be time independent for $v=v_\varphi(\gamma)$.
By contrast, the {\it group velocity} is defined by the saddle point 
$\gamma_c$
of the exponential
phase factor
\begin{equation}
v=\left.\frac{d\omega}{d\gamma}\right|_{{\gamma_c}}\equiv v_g\ .
\label{groupvelocity}
\end{equation} 

Let us show that the actual velocity $v$ of the front
depends on the competition between the singularities of
$u_0(\gamma)$ and 
the saddle point of the phase 
factor.

In Eq.(\ref{inicond}),
$u_0(\gamma)$ is the initial condition in Mellin space,
{\it i.e.} the Mellin
transform of $u(x,t\!=\!0)$.
The physically appropriate 
assumption is that $u_0(\gamma)$ is a monotoneous
function smoothly connecting 1 to 0 as $x$ goes from $-\infty$ to $+\infty$,
with asymptotic behavior $u(x,t\!=\!0)\sim e^{-\gamma_0 x}$ 
\cite{Munier:2003vc}.
Then $u_0(\gamma)$ has singularities on 
the real axis starting at $0$ and extending
to the left, and at $\gamma_0$ to the right.
A simple case study is $u(0,x)=1$ for $x\leq 0$
and $u(0,x)=e ^{-\gamma_0 x}$ for $x>0$, whose exact Mellin transform
is $u_0(\gamma)=1/\gamma+1/(\gamma_0-\gamma)$. The contour ${\cal C}$
in~(\ref{inicond}) runs
parallel to the imaginary axis and crosses the real axis
at a point between 0 and $\gamma_0$.  
There are three relevant cases
(see Fig.1):

\noindent {\it (i)} $\gamma_0<\gamma_c$: dominance of initial conditions.
The integral~(\ref{inicond}) 
defining $u$ is dominated by the pole at $\gamma_0$, thus
$u\sim\exp\{-\gamma_0(x_{\text{\tiny WF}}+vt)+\omega(\gamma_0)t\}$
and the velocity of the front is determined
by consistency to be $v\equiv \omega(\gamma_0)/\gamma_0$, which ensures the
time-independence in the reference frame of the front.

\noindent {\it (ii)} $\gamma_0>\gamma_c$: dominance of the saddle point.
The integral in Eq.(\ref{inicond}) is now
 $\sim\exp\{-\gamma_c(x_{\text{\tiny WF}}+vt)+\omega(\gamma_c)t\}$
and thus the velocity $v$ of the front is defined by
\begin{equation}
v=v_g=\frac{\omega(\gamma_c)}{\gamma_c}
=\min_\gamma\frac{\omega(\gamma)}{\gamma}\ .
\label{frontvelocity}
\end{equation}
Indeed, in this case, the group velocity is identical to
the minimum of the phase velocity\footnote{Note that the relation 
$v_g=v_{\varphi},$ in analogy with wave physics, has been written in 
Ref.\cite{GLR}.}.
In the previous case instead, the front velocity is 
$v=\omega(\gamma_0)/\gamma_0>v_g$.

\noindent {\it (iii)} $\gamma_0=\gamma_c$: critical case.
Both the pole and the saddle point contribute, and $v=v_g$.

Now that we have determined the velocity of the wave front, 
reporting it in Eq.(\ref{inicond}) and performing the integral in the
large $t$ limit gives the
asymptotic expression for the form of the front:
\begin{equation}
u(x,t)
\underset{t\rightarrow\infty}{\sim}\begin{cases}
\ \ e^{-\gamma_0 x_{\text{\tiny WF}}}& \text{if $\gamma_0<\gamma_c$}\\
\ \ e^{-\gamma_c x_{\text{\tiny WF}}}& \text{if $\gamma_0\geq \gamma_c$}
\end{cases}\ \ \ .
\label{asymptotic}
\end{equation}
The typical wave front solutions
are illustrated on Fig.2. 
The upper plot displays a typical feature of the wave front development
for the case $\gamma_0<\gamma_c$. Here the front moves with
the velocity $v_->v_g$ and keeps the memory of the initial condition.
The lower plot illustrates the case $\gamma_0>\gamma_c$. In this case,
the wave front develops with the fixed group velocity $v_g$.
The forward part of the curve (after the knee) 
is the part of the front which keeps the memory of the initial
condition, which is a typical feature of a ``pulled'' 
front \cite{vansaarloos}.
The middle plot is for the critical case $\gamma_0=\gamma_c$.

In the following, we
will focus on the case in which the initial condition is
steep enough, so that the velocity of the front is the group 
velocity~(\ref{groupvelocity}) and the form of the wave front is the second
equation of~(\ref{asymptotic}).

\begin{figure}[ht]
\begin{center}
\epsfig{file=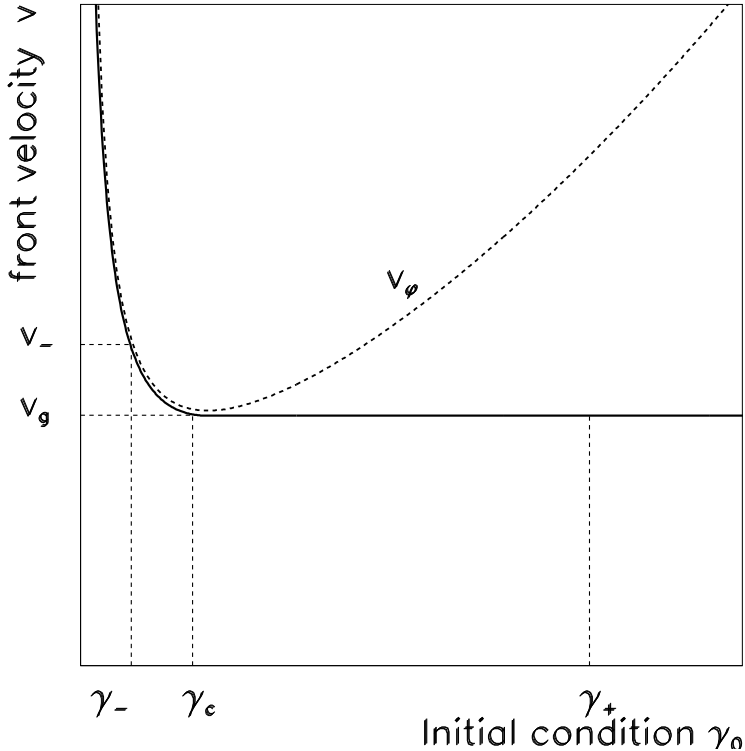,width=11cm}
\end{center}
\caption{The front velocity as a function of the steepness 
$\gamma_0$ of the
  initial condition at $x\rightarrow\infty$ 
($u(x,t\!=\!0)\sim e ^{-\gamma_0 x}$). 
{\it Continuous line}: the actual
  front 
velocity. {\it Dashed line}: the phase velocity
for a wave with wave number $\gamma_0$.
The abscissa points 
$\gamma_-$, $\gamma_+$, $\gamma_c$ correspond to initial conditions resp.
  below, larger than and equal to the critical value (resp. cases {\it (i)},
{\it (ii)} and {\it (iii)}, see text). } 
\end{figure}

\begin{figure}[ht]
\begin{center}
\epsfig{file=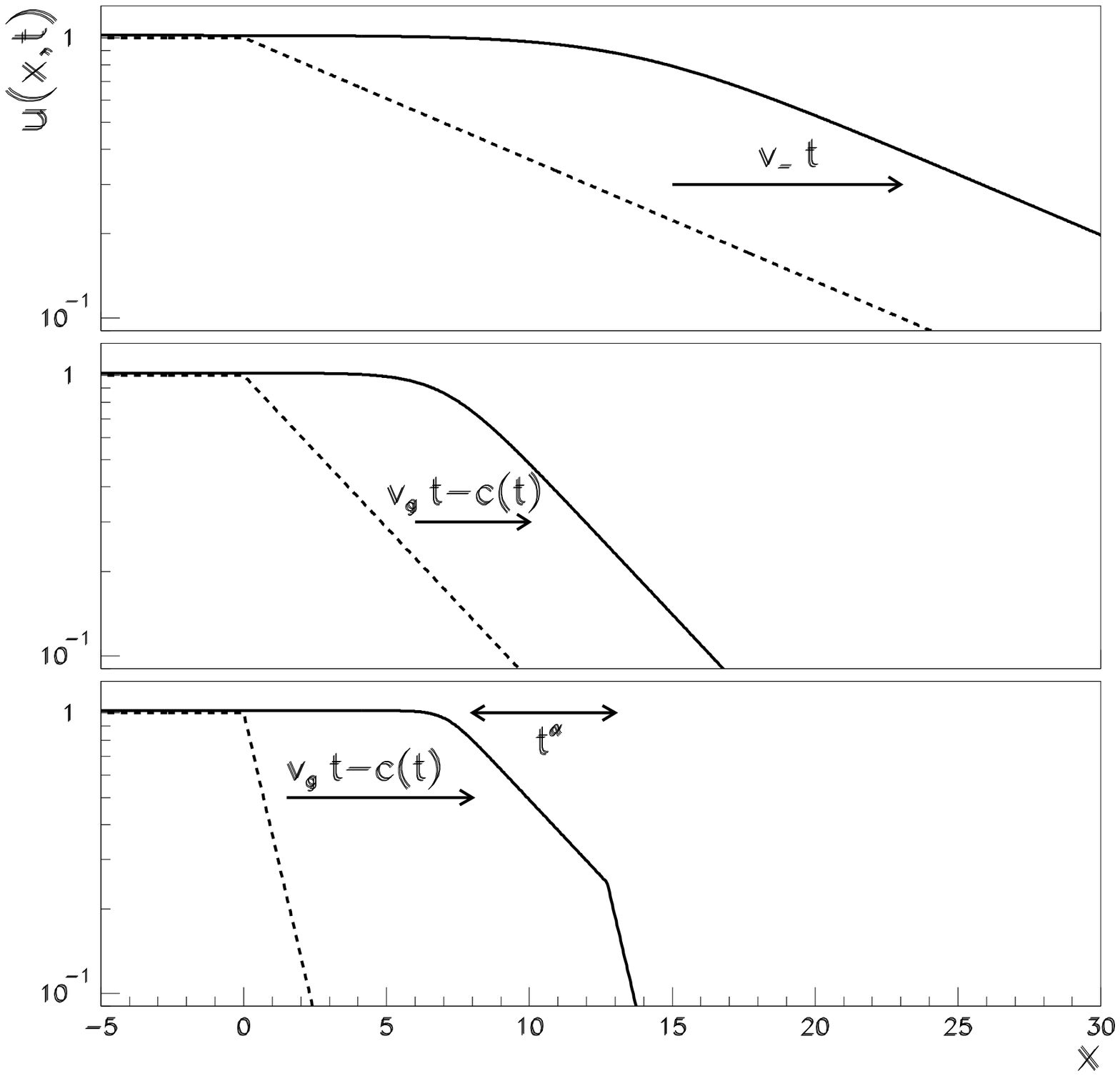,width=13cm}
\end{center}
\caption{Picture of the
evolution of a front for three typical initial conditions. {\it Dashed
  line}: the initial profile $u(x,t=0)$. {\it Full line}: the evolved front
  after at large time $t$. {\it Upper
  plot} (case {\it(i)}): 
the exponential slope $\gamma_0$ of the initial condition is
  $\gamma_-<\gamma_c$. The front propagates at the phase
  velocity $v_-=v_\varphi(\gamma_-)$. 
{\it Middle plot} (case {\it (iii)}): 
$\gamma_0=\gamma_c$. The front propagates at velocity $v_g$.
{\it Lower plot} (case {\it (ii)}): 
$\gamma_0=\gamma_+>\gamma_c$. The front still propagates
  at velocity $v_g$, the size of the front is finite of order
  $t^\alpha$, and its slope is given by $\gamma_c$.}
\end{figure}


It is interesting for our purpose to go beyond the 
asymptotics~(\ref{asymptotic})
and study how the front reaches its asymptotic form as a function of time. 
For this sake, we shall use the
following ansatz in the neighbourhood of the front:
\begin{equation}
u(x,t)\underset{t\ \text{large}}{=}
t^\alpha\, G\left(\frac{x_{\text{\tiny WF}}+c(t)}{t^\alpha}\right)
e ^{-\gamma_c(x_{\text{\tiny WF}}+c(t))}\ ,
\label{ansatz}
\end{equation}
where $c(t)$ is a subdominant
function of time  (with respect to $v_gt$).
Indeed, the front velocity receives some subasymptotic 
corrections  w.r.t. the constant asymptotic 
velocity $v_g$, contained in the derivative 
$\dot c(t)=dc(t)/dt$.
The factors $t^\alpha$ describe a
diffusion-like evolution which will be made explicit in each particular
case.
Hence the function
$t^\alpha G$ represents a subasymptotic 
correction to the form of the front~(\ref{asymptotic})
at large times.
This ansatz was proven to be relevant for 
KPP-like equations~\cite{Brunet:1997xx}.
Its key property is to be a solution of the linearized equation
at large $t$ in the neighbourhood of the front, 
for some $\alpha$ and $c(t)$ to be determined. 
Note that 
since $u(x,t)$ only depends on 
${x_{\text{\tiny WF}}+c(t)}$
when the front has 
reached its asymptotic form of
a pure traveling wave, 
$G(z)$ is required to
vanish like $z$ for $z\rightarrow 0$.

The features of this ansatz are displayed on the lowest plot in
Fig.2. One sees that a steep initial condition characterized by the
exponential decay slope $\gamma_0$ evolves, after a given
time $t$, into a front characterized by the
critical exponential decay slope $\gamma_c$. 
Its width is of order $t^\alpha$ as we will show later on.\\


\section{BK equation with fixed QCD coupling}
\label{3}

Let us first
apply the formalism exposed above to 
the BK equation with fixed coupling $\bar\alpha$.
The linear part of Eq.(\ref{bk}) is solved by
\begin{equation}
{\cal N}(k,Y)=\int\frac{d\gamma}{2i\pi}
{\cal N}_0(\gamma)
\exp\left(-\gamma L
+\bar\alpha\chi(\gamma)Y\right)\ .
\label{linsol}
\end{equation}
The rapidity $Y$ is the evolution variable, and thus plays the
r{\^o}le of the time $t$, while 
$L=\log(k^2/\Lambda_{\text{\footnotesize QCD}}^2)$ plays the r{\^o}le of the
spatial coordinate $x$.

Let us discuss the relevant initial condition ${\cal N}_0(\gamma)$ for
QCD. The essential property for the derivation is
that at large $L$ ({\it i.e.} large $k$), 
${\cal N}(k,Y\!=\!0)\sim k^{-2}$: this
corresponds to the color transparency property of gauge
theories, and yields $\gamma_0\!=\!1$.
For $L$ small or negative, the dynamics is unknown since it belongs
to the nonperturbative regime of QCD. However, interestingly enough, 
it does not play a direct r{\^o}le 
in the final result
for the front formation. One may assume that ${\cal N}(k,Y\!=\!0)$ 
is bounded \cite{Munier:2003vc}, 
which is a way to impose unitarity on the
initial condition. In Mellin space, this leads to singularities
of ${\cal N}_0(\gamma)$ 
distributed on the intervals $]-\!\infty,0]$ and $[1,+\!\infty[$.
We shall verify in a moment that the 
wave number $\gamma_c$  which minimizes 
the {\it phase velocity}, {\it i.e.} the
wave number of the partial wave which moves with
the {\it group velocity}, is between 0 and 1. Thus we are in the case
{\it (ii)} of Sec.\ref{2} in which the saddle point dominates.

We shall now proceed to the determination of the analogous of the group
velocity $v_g$.
By comparison of Eq.(\ref{linsol}) to Eq.(\ref{inicond}), one sees that
the dispersion relation reads
\begin{equation}
\omega(\gamma)=\bar\alpha\chi(\gamma)\ .
\end{equation}
The phase velocity
$v_\varphi(\gamma)=\omega(\gamma)/\gamma$ 
has a minimum at $\gamma_c$. From
Eqs.(\ref{groupvelocity},\ref{frontvelocity}),
$\gamma_c$ solves the implicit equation
\begin{equation}
\gamma_c\chi ^\prime(\gamma_c)=\chi(\gamma_c)\ ,
\label{defbargamma}
\end{equation}
which yields $\gamma_c=0.6275...$~. This number indeed stands between
0 and 1, and thus we are in a similar situation as the one depicted on the
third plot of Fig.2: 
the front velocity 
is blocked at the value
\begin{equation}
v_g=\bar\alpha\,\frac{\chi(\gamma_c)}{\gamma_c}\ .
\label{fcvelocity}
\end{equation}

To study the formation of the front,
we will stick to the ``diffusive'' approximation
of the BK equation~(\ref{bk}), defined by
keeping terms up to the second order
only in the expansion~(\ref{chiexpansion}):
\begin{equation}
\chi(-\partial_L)=\chi(\gamma_c)+(-\partial_L-\gamma_c{\mathbf 1})
\chi ^\prime(\gamma_c)+{\scriptstyle\frac12}
(-\partial_L-\gamma_c{\mathbf 1})^2\chi ^{\prime\prime}(\gamma_c)\ .
\end{equation}
Within this approximation, the nonlinear equation becomes
\begin{equation}
{\partial_Y}{\cal N}=-v_g\partial_L{\cal N}
+{\scriptstyle\frac12}\bar\alpha{\ccc}(\partial_L+\gamma_c{\bf 1})^2{\cal
  N}-\bar\alpha{\cal N}^2\ ,
\label{bkapprox}
\end{equation}
which is equivalent to the KPP equation \cite{Munier:2003vc}.

We now use the ansatz~(\ref{ansatz}) as the appropriate
solution of the linearized version of Eq.(\ref{bkapprox}) near the
wave front.
Setting  $u\!=\!{\cal N}$, 
$x\!=\!L$ and $t\!=\!Y$,
the linear part of
Eq.(\ref{bkapprox}) turns into an ordinary
differential equation for $G(z)$, namely
\begin{equation}
{\scriptstyle\frac12}{\bar\alpha \ccc}Y^{-\alpha} G^{\prime\prime}(z)
+(\alpha z Y^{\alpha-1}-\dot c(Y))G^\prime(z)+Y^{\alpha-1}(\gamma_c\,\dot c(Y)\,
Y-\alpha)G(z)=0
\label{bkapproxlineaire}
\end{equation}
where $z$ is related to the physical variables through 
\begin{equation}
z\equiv\frac{x_{\text{\tiny WF}}+c(t)}{t^\alpha}=
\frac{(L-v_g Y)+c(Y)}{Y^{\alpha}}\ .
\end{equation}
The different
terms in Eq.(\ref{bkapproxlineaire}) 
contribute equally if $\alpha={\scriptstyle\frac12}$ and
$\dot c(Y)=\beta/Y$, where $\beta$ is a constant. 
This value of $\alpha$ ensures that the terms proportional to
$Y^{-\alpha}$ and $Y^{\alpha-1}$ contribute to the same order in $Y$.
Using these values,
Eq.(\ref{bkapproxlineaire}) boils down
to a simple second order differential equation
\begin{equation}
{\scriptstyle\frac12}{\bar\alpha \ccc} G^{\prime\prime}(z)
+{\scriptstyle\frac12} 
z G^\prime(z)+(\beta\gamma_c-{\scriptstyle\frac12})G(z)=0\ .
\end{equation}
The constant $\beta$ is determined by requiring that $G$ behave smoothly
for $z\rightarrow +\infty$, which fixes uniquely%
\footnote{The full asymptotics of $G$ in the framework
of the KPP equation can be found in Ref.\cite{Brunet:1997xx}.}
$\beta=3/(2\gamma_c)$ \cite{Brunet:1997xx}. 
The solution which vanishes at $z=0$ is unique up to a constant and reads
\begin{equation}
G(z)=\text{const.}\times\sqrt{\frac{2}{\bar\alpha \ccc}}\,z\,\exp
\left(-\frac{z^2}{2\bar\alpha\ccc}\right)\ .
\end{equation}
In terms of physical variables, the gluon density near the wave
front thus reads
\begin{equation}
{\cal N}(k/Q_s(Y),Y)=\text{const.}\times
\,\sqrt{\frac{2}{\bar\alpha\chi ^{\prime\prime}(\gamma_c)}}
\log\left(\frac{k^2}{Q_s^2(Y)}\right)
\left(
\frac{k^2}{Q_s^2(Y)}\right)^{-\gamma_c}
\exp\left(-\frac{1}{2\bar\alpha\chi ^{\prime\prime}(\gamma_c)Y}
\log^2\left(\frac{k^2}{Q_s^2(Y)}\right)\right)
\label{nfixedalpha}
\end{equation}
with the saturation scale defined as\footnote{Note that~(\ref{qsfixedalpha}) 
reproduces the result of~\cite{Munier:2003vc}
after expansion of $\gamma_c$ around ${\scriptstyle \frac12}$.} 
\begin{equation}
Q_s^2(Y)=Q_0 ^2
\exp\left(\bar\alpha\frac{\chi(\gamma_c)}{\gamma_c}Y
-\frac{3}{2\gamma_c}\log Y\right)\ .
\label{qsfixedalpha}
\end{equation}
$Q_0$ absorbs undetermined constants 
but remains of order $\Lambda_{\text{\footnotesize QCD}}$.\\


\section{BK equation with running QCD coupling}
\label{4}

We turn to the case in which
 the coupling in Eq.(\ref{bk}) runs with $L$ as given 
by Eq.(\ref{baralpharun}).
The BK equation reads
\begin{equation}
bL\ {\partial_Y}{\cal N}=
\chi\left(-\partial_L\right){\cal N}
-{\cal N}^2\ ,
\end{equation}
with $b$ defined in Eq.(\ref{baralpharun}).
Using a double Mellin transform, the 
linearized version of the equation gives the solution
\begin{equation}
{\cal N}(k,Y)=\int\frac{d\gamma}{2i\pi}\int\frac{d\omega}{2i\pi}
{\cal N}_0(\gamma,\omega)
\exp\left(-\gamma L+\omega Y+\frac{1}{b\omega}X(\gamma)\right)\ ,
\end{equation}
where 
\begin{equation}
X(\gamma)=\int_{\widehat\gamma}^\gamma d\gamma^\prime\,\chi(\gamma^\prime)\ .
\label{X}
\end{equation}
At this stage, $\widehat\gamma$ is an undetermined
constant. The initial condition 
${\cal N}_0(\gamma,\omega)$ depends now both on $\omega$ and $\gamma$, 
but has to
satisfy the same requirements as in the fixed coupling case 
(color transparency at large $k$, saturation at smaller $k$) resulting in
a specific analytic structure in $\gamma$ as already
discussed in Sec.III.

Performing the integral over $\omega$ in the large $Y$ limit,
the integral is dominated by
a saddle point at
$\omega_c=\sqrt{bX(\gamma)/Y}$.
The solution reads 
\begin{equation}
{\cal N}(L,Y)=\int\frac{d\gamma}{2i\pi}{\cal N}_0(\gamma)
\exp\left(-\gamma L+\sqrt{Y}\,\sqrt{\frac{4X(\gamma)}{b}}
\right)\ ,
\end{equation}
where ${\cal N}_0(\gamma)$ keeps track of all prefactors.
Now interpreting $\sqrt{Y}$ as the time variable $t$, 
the gluon density turns out to be also a linear
superposition of traveling waves labelled by $\gamma$,
and thus can be treated by the general method exposed in Sec.\ref{2}.
In this case,
the asymptotic 
dispersion relation is
\begin{equation}
\omega(\gamma)\equiv\sqrt{\frac{4X(\gamma)}{b}}\ .
\end{equation}
The minimum $\bar\gamma$ of
$v_\varphi(\gamma)=\omega(\gamma)/\gamma$ 
satisfies the equation
\begin{equation}
2X(\bar\gamma)={\bar\gamma}\,\chi(\bar\gamma)\ ,
\end{equation}
still depending at this stage on the choice of $\widehat\gamma$ 
in Eq.(\ref{X}).
The velocity of the front (see Eq.(\ref{groupvelocity}))
is given by
\begin{equation}
v_g
=\sqrt{\frac{2\chi(\bar\gamma)}{b\bar\gamma}}\ .
\end{equation}
Requiring the velocity $v_g$ to become independent of the choice of
$\widehat\gamma$ leads to the constraint
$dv_g(\widehat\gamma)/d\widehat\gamma=0=dv_g(\bar\gamma)/d\bar\gamma$, namely
\begin{equation}
\bar\gamma\,\chi^\prime(\bar\gamma)=
{\chi(\bar\gamma)}\ .
\end{equation}
The comparison with Eq.(\ref{defbargamma}) shows that 
$\bar\gamma$ is identical to $\gamma_c=0.6275...$, the critical wave number in 
the
non-running case, and thus the velocity of the front reads
\begin{equation}
v_g=v(\gamma_c)=\sqrt{\frac{2\chi(\gamma_c)}{b\gamma_c}}\ .
\end{equation}

Once the asymptotic velocity is known, we compute the transition to
saturation through subleading corrections to $v_g$.
To this aim, one expands the linearized BK 
equation around $\gamma_c$ to get
\begin{equation}
\frac{bL}{2t}{\partial_t {\cal N}}
=-\frac{b v_g^2}{2}\partial_L {\cal N}
+\frac{\ccc}{2}(\partial_L^2 {\cal N}+2\gamma_c\partial_L {\cal N}
+\gamma_c ^2 {\cal N})\ .
\label{bksimplif2}
\end{equation}
One uses the ansatz~(\ref{ansatz}) for the solution to this 
equation at large $t$, which turns it
into an ordinary second order differential equation. 
After tedious but straightforward compilation of all terms, 
one determines 
the coefficient $\alpha$ and the function
$c(t)$ by requiring that they all contribute equally.
One finds $\alpha={\scriptstyle \frac13}$ and $\dot c(t)=\beta
t^{-2/3}$, where $\beta$ still needs to be fixed.
The value of $\alpha$ comes from keeping relevant terms
in $Y^{-\alpha}$ and $Y^{2\alpha-1}$, which all contribute at large $Y$.

Finally, Eq.(\ref{bksimplif2}) reduces to the Airy equation
\begin{equation}
G^{\prime\prime}(z)=
\frac{b\gamma_c v_g}{\ccc}(z-4\beta)
\,G(z)\ .
\end{equation}
There are two independent solutions to this equation, {\it e.g.} 
Ai and Bi.
Imposing the regularity at $z=+\infty$ selects
\begin{equation}
G(z)=\text{const.}\times
\text{Ai}\left(\left(\frac{\gamma_c v_g b}{\ccc}\right)^{1/3}
(z-4\beta)\right)\ .
\end{equation}
The condition $G(z)\sim z$ for small $z$ is satisfied if
\begin{equation}
\beta=-\frac14\left(
\frac{\ccc}{\gamma_c v_g b}
\right)^{1/3}\xi_1\ ,
\end{equation}
where $\xi_1=-2.338...$ is the rightmost zero of the Airy function.
The complete result reads, in physical variables
\begin{equation}
{\cal N}(k/Q_s(Y),Y)=
\text{const.}\times Y^{1/6}\left(\frac{k^2}{Q_s^2(Y)}\right)^{-\gamma_c}
\text{Ai}\left(\xi_1+
\left(\frac{\sqrt{2b\gamma_c\chi(\gamma_c)}}
{\chi ^{\prime\prime}(\gamma_c)}\right)^{1/3}
\frac{\log (k^2/Q_s^2(Y))}{Y^{1/6}}\right)\ ,
\label{nrunningalpha}
\end{equation}
where the saturation scale is given by
\begin{equation}
Q_s^2(Y)=\Lambda_{\text{\footnotesize QCD}} ^2
\exp\left(\sqrt{\frac{2\chi(\gamma_c)}{b\gamma_c}Y}
+\frac34\left(\frac{\chi ^{\prime\prime}(\gamma_c)}
{\sqrt{2b\gamma_c\chi(\gamma_c)}}
\right)^{1/3}\!\xi_1 Y^{1/6}\right)\ ,
\label{qsrunningalpha}
\end{equation}
up to a multiplicative constant.

\begin{figure}[ht]
\begin{center}
\epsfig{file=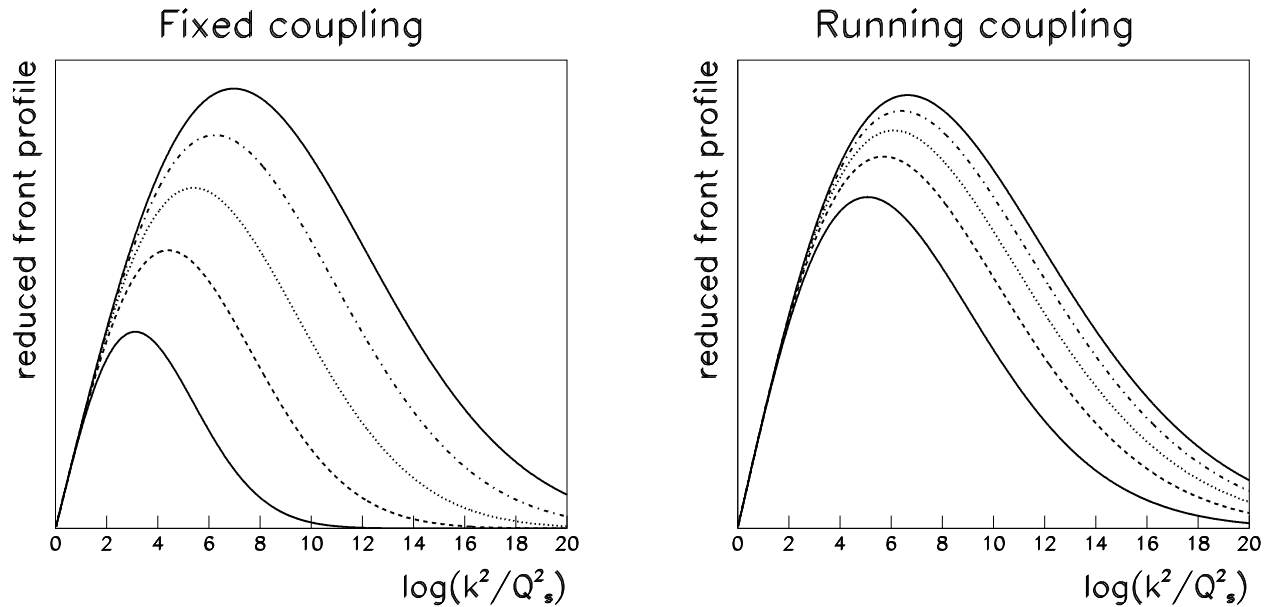,width=17cm}
\caption{Evolution of the reduced front profile in the case 
of fixed coupling (left) and running coupling (right). 
The reduced front profile 
$(k^2/Q_s^2)^{\gamma_c}\,{\cal N}(k/Q_s(Y),Y)$ is plotted against
$\log(k^2/Q_s^2)$ for different rapidities.
 The various lines correspond to rapidities from 2 (lower
  curves, full line) up to 10 (upper curves).
Note the similarity of the wave fronts, but the quicker time evolution
(in $\sqrt{t}$) for fixed coupling, by contrast with the slow time
evolution (in $t^{1/3}$) for the running coupling case.}
\end{center}
\end{figure}


\section{Discussion and conclusion}
\label{5}

Let us discuss our final results for fixed
(\ref{nfixedalpha},\ref{qsfixedalpha}) and running
(\ref{nrunningalpha},\ref{qsrunningalpha}) coupling.
In both cases, we get the leading exponential terms
(resp. $\bar\alpha(\chi(\gamma_c)/\gamma_c)Y$ and
$\sqrt{2(\chi(\gamma_c)/b\gamma_c)Y}$) in the
expressions for the saturation scale found 
in Refs.\cite{Iancu:2002tr,Mueller:2002zm}.
These terms
are supplemented by subleading corrections
in rapidity, which are precisely those present in Ref.\cite{Mueller:2002zm},
and which are respectively proportional to 
$\log Y$ and $Y^{1/6}$.
They are related to the slowing down of the
velocity due to formation of the wave front.
We expect this slowing down to be a very general feature of the
transition to saturation. The expressions for the front profiles 
exhibit geometric scaling
violations which are 
due to the time ({\it i.e.} rapidity) delay for the
formation of the front.

Comparing to the results which were obtained from the linear evolution
equation supplemented by absorptive boundary conditions
\cite{Mueller:2002zm}, we have confirmed these results for the running
coupling case\footnote{
Note that an extension to the (resummed) next-to-leading order BFKL kernel
has been found in Ref.\cite{Triantafyllopoulos:2002nz}. It will be interesting 
to see how this result can also be obtained by our method.}, up to a constant 
multiplicative factor in the
definition of the saturation scale $Q_s$. For the fixed coupling case,
the same is true within the geometric scaling region $k\sim
Q_s(Y)$. Our formula~(\ref{nfixedalpha}) provides a hint on the
geometric scaling violations in this case also. A stimulating  and fruitful 
challenge for future applications is to understand why our method and the one of 
Ref.\cite{Mueller:2002zm} appear to be equivalent for the treated problems. It 
is interesting to note that numerical simulations of the fixed coupling case 
\cite{armesto} and of the full equation with running coupling \cite{weigert} 
confirm the validity of the analytic estimates.

To summarize,
in this paper, we have proposed a general method, physically
appealing, which allows to
determine both the saturation scale and the transition of the gluon
density towards saturation for the solution of the Balitsky-Kovchegov
nonlinear equation. The method is based on a direct analogy with
traveling wave solutions of nonlinear equations of the KPP type.
The saturation scale is related to the
velocity of the wave front, and the transition to saturation of the
gluon distribution to the diffusive type formation of the front.
In particular, we have studied directly
the solution to the nonlinear equation,
instead of assuming specific boundary conditions on the linear one.

As first applications of this method, 
using a general ansatz \cite{Brunet:1997xx},
we have derived in a rather simple way 
the physical solutions to the BK equation at large rapidity
for both fixed and running coupling, see 
Eqs.(\ref{nfixedalpha},\ref{qsfixedalpha}) 
and~(\ref{nrunningalpha},\ref{qsrunningalpha}).
The approach to saturation (leading
to geometric scaling violations) corresponds to ordinary diffusion 
(namely with a diffusion length going like $\sqrt{t}$) in the fixed coupling 
case
and to anomalous
diffusion (in $t^{1/3}$) in the running coupling case.
These contrasted diffusion patterns towards saturation are made explicit in the
evolution of the reduced front profile 
$(k^2/Q_s^2)^{\gamma_c}\,{\cal N}(k/Q_s(Y),Y)$
shown on Fig.3.

We think that the method that we have presented is general enough
to allow for desirable extensions to the full content of saturation
equations.

\begin{acknowledgments}
We thank Bernard Derrida for his illuminating
advices on traveling waves and KPP-like equations and Dionysis 
Triantafyllopoulos for his comments.
\end{acknowledgments}


\begin{thebibliography}{10}


\bibitem{GLR}  L.V. Gribov, E.M. Levin and M.G. Ryskin, {Phys. Rep.} {\bf 
100}, 1 (1983).


\bibitem{qiu} 
{A.H.\ Mueller, J.\ Qiu}, { Nucl.\ Phys.} {\bf B268}, 427 {(1986)}.




 \bibitem{venugopalan}
  {L. McLerran and R. Venugopalan}, {Phys.\ Rev.} {\bf D49}, 2233 (1994);
{\it ibid.}, 3352 { (1994)};
{\it ibid.}, {\bf D50}, 2225{ (1994)};
  {A. Kovner, L. McLerran and H. Weigert}, {Phys.\ Rev.} {\bf D52}, 6231{
    (1995)}~; {\it ibid.}, 3809{ (1995)};
   {R. Venugopalan}, {Acta Phys.Polon.} {\bf B30}, 3731 (1999);
E.~Iancu, A.~Leonidov, and L.~McLerran,
 { Nucl. Phys.}~{\bf A692}, 583 (2001); {\it idem},
{ Phys. Lett.} {\bf B510}, 133 (2001);
E.~Iancu and L.~McLerran, { Phys. Lett.} {\bf B510}, 145 (2001);
E.~Ferreiro, E.~Iancu, A.~Leonidov and L.~McLerran, 
{ Nucl. Phys.} {\bf A703}, 489 (2002);
H.~Weigert, { Nucl. Phys.} {\bf A703}, 823 (2002).


\bibitem{Balitsky:1995ub}
I.~Balitsky,
Nucl.\ Phys.\ {\bf B463}, 99 (1996);
Y.~V. Kovchegov,
\newblock Phys. Rev. {\bf D60}, 034008 (1999);
\newblock
{\it ibid.},
{\bf D61}, 074018 (2000).

\bibitem{levin} 
{E.\ Levin, J.\ Bartels}, { Nucl.\ Phys.} {\bf B387}, 617 {(1992)};
Y.~V. Kovchegov,
\newblock Phys. Rev. 
{\bf D61}, 074018 (2000).
{E.\ Levin, K.\ Tuchin}, { Nucl.\ Phys.} {\bf A693}, 787 {(2001)}; 
{\it ibid.},
{\bf A691}, 779 {(2001)}.
.

\bibitem{Mueller:2002zm}
A.~H. Mueller and D.~N. Triantafyllopoulos,
\newblock Nucl. Phys. {\bf B640}, 331 (2002).

\bibitem{BFKL}
L.~N. Lipatov,
\newblock Sov. J. Nucl. Phys. {\bf 23}, 338 (1976);
E.~A. Kuraev, L.~N. Lipatov, and V.~S. Fadin,
\newblock Sov. Phys. JETP {\bf 45}, 199 (1977);
I.~I. Balitsky and L.~N. Lipatov,
\newblock Sov. J. Nucl. Phys. {\bf 28}, 822 (1978).

\bibitem{Munier:2003vc}
S.~Munier and R.~Peschanski,
{\it ``Geometric scaling as traveling waves''},
to appear in Phys. Rev. Lett.,
[arXiv:hep-ph/0309177].

\bibitem{Stasto:2000er}
A.~M. Sta\'sto, K.~Golec-Biernat, and J.~Kwiecinski,
\newblock Phys. Rev. Lett. {\bf 86}, 596 (2001).

\bibitem{vansaarloos}
Wim van Saarloos,
\newblock Phys. Rep. {\bf 386}, 29 (2003).

\bibitem{Fisher:37}
R.~A. Fisher,
\newblock Ann. Eugenics {\bf 7}, 355 (1937);
A.~Kolmogorov, I.~Petrovsky, and N.~Piscounov,
\newblock Moscou Univ. Bull. Math. {\bf A1}, 1 (1937).


\bibitem{Bramson}
M.~Bramson,
\newblock Memoirs of the American Mathematical Society {\bf 285} (1983).

\bibitem{Brunet:1997xx}
E.~Brunet and B.~Derrida,
\newblock Phys. Rev. {\bf E56}, 2597 (1997) and
references therein.


\bibitem{Iancu:2002tr}
E.~Iancu, K.~Itakura and L.~McLerran,
Nucl.\ Phys.\ A {\bf 708}, 327 (2002).





\bibitem{Triantafyllopoulos:2002nz}
D.~N.~Triantafyllopoulos,
Nucl.\ Phys.\ {\bf B648}, 293 (2003).

\bibitem{armesto} J.L. Albacete, N. Armesto, A.Kovner, C.A. Salgado, U.A. 
Wiedemann, {\it ``Energy Dependence of the Cronin Effect from Non-Linear QCD 
Evolution''}, [arXiv:hep-ph/0307179].


\bibitem{weigert} K. Rummukainen, H. Weigert,
{\it ``Universal features of JIMWLK and BK evolution at small x''}, 
[arXiv:hep-ph/0309306].

\end{thebibliography}
\end{document}